\documentclass{article}

\usepackage{arxiv}

\usepackage[utf8]{inputenc} 
\usepackage[T1]{fontenc}    
\usepackage{hyperref}       
\usepackage{url}            
\usepackage{booktabs}       
\usepackage{amsfonts}       
\usepackage{nicefrac}       
\usepackage{microtype}      
\usepackage{lipsum}		
\usepackage{graphicx}
\usepackage{natbib}
\usepackage{doi}

\usepackage{bibunits} 

\setcitestyle{super} 

\usepackage{url}
\usepackage{xspace}
\usepackage{comment}
\usepackage{amsmath}

\newcommand{\figref}[1]{Figure~\ref{fig:#1}}

\newcommand{\tableref}[1]{Table~\ref{tab:#1}}
\newcommand{\suppsecref}[1]{Supplementary~\ref{supp:#1}}

\newcommand{\degree}{$^{\circ}$}

\newcommand{\cdms}{\,cd/m$^2$\xspace}
\newcommand{\cpd}{\,cpd\xspace}
\newcommand{\ppd}{\,ppd\xspace}
\newcommand{\cm}{\,cm\xspace}

\renewcommand{\eqref}[1]{Eq.~\ref{eq:#1}}

\newcommand{\code}[1]{\texttt{#1}}

\usepackage{placeins}

\usepackage{subcaption} 

\usepackage{multirow}
\usepackage[table]{xcolor}

\newcommand{\Sconscol}[1]{\ensuremath{{S}_{0}^\text{#1}}} 
\newcommand{\krhocol}[1]{\ensuremath{{k}_{\rho}^\text{#1}}} 
\newcommand{\kecccol}[1]{\ensuremath{{k}_{e}^\text{#1}}} 

\newcommand{\Scons}[0]{\ensuremath{{S}_{0}}} 
\newcommand{\krho}[0]{\ensuremath{{k}_{\rho}}} 
\newcommand{\kecc}[0]{\ensuremath{{k}_{e}}} 

\newcommand{\meangauss}[0]{\ensuremath{\mu_{e}}}
\newcommand{\siggauss}[0]{\ensuremath{\sigma_{e}}}
\newcommand{\scgauss}[0]{\ensuremath{A_{e}}}

\newcommand{\snellfrac}[0]{\ensuremath{r_{\text{snellen}}}}
\newcommand{\dchart}[0]{\ensuremath{d_{\text{chart}}}}
\newcommand{\dnorm}[0]{\ensuremath{d_{\text{norm}}}}
\newcommand{\dwidth}[0]{\ensuremath{d_{\text{width}}}}
\newcommand{\rw}[0]{\ensuremath{r_{\text{w}}}}
\newcommand{\dv}[0]{\ensuremath{d_{\text{v}}}}

\title{Resolution limit of the eye: how many pixels can we see?}

\author{ \href{https://orcid.org/0000-0002-8142-5611}{\includegraphics[scale=0.06]{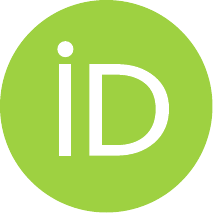}\hspace{1mm}Maliha~Ashraf}\\
	Department of Computer Science \& Technology\\
	University of Cambridge\\
	Cambridge, CB3 0FD, UK\\
	\texttt{ma905@cam.ac.uk} \\
	\And
	\href{https://orcid.org/0000-0002-7367-0131}{\includegraphics[scale=0.06]{orcid.pdf}\hspace{1mm}Alexandre~Chapiro} \\
	Meta\\
	Applied Perception Science\\
	Sunnyvale, CA 94089, USA\\
	\texttt{achapiro@meta.com} \\
	\And
	\href{https://orcid.org/0000-0003-2353-0349}{\includegraphics[scale=0.06]{orcid.pdf}\hspace{1mm}Rafa{\l} K.~Mantiuk} \\
	Department of Computer Science \& Technology\\
	University of Cambridge\\
	Cambridge, CB3 0FD, UK\\
	\texttt{rafal.mantiuk@cl.cam.ac.uk} \\
}

\renewcommand{\shorttitle}{Resolution limit of the eye}

\hypersetup{
pdftitle={Resolution limit of the eye: how many pixels can we see?},
pdfauthor={Maliha~Ashraf, Alexandre~Chapiro, Rafa{\l} K.~Mantiuk},
pdfkeywords={visual resolution, retinal resolution, display resolution, human visual acuity},
}

\begin{document}
\maketitle

\begin{abstract}

 As large engineering efforts go towards improving the resolution of mobile, AR and VR displays, it is important to know the maximum resolution at which further improvements bring no noticeable benefit. This limit is often referred to as the \emph{retinal resolution}, although the limiting factor may not necessarily be attributed to the retina. 
To determine the ultimate resolution at which an image appears sharp to our eyes with no perceivable blur, we created an experimental setup with a sliding display, which allows for continuous control of the resolution. The lack of such control was the main limitation of the previous studies. 
We measure achromatic (black-white) and chromatic (red-green and yellow-violet) resolution limits for foveal vision, and at two eccentricities (10 and 20 deg). 
Our results demonstrate that the resolution limit is higher than what was previously believed, reaching 94 pixels-per-degree (ppd) for foveal achromatic vision, 89 ppd for red-green patterns, and 53 ppd for yellow-violet patterns. We also observe a much larger drop in the resolution limit for chromatic patterns (red-green and yellow-violet) than for achromatic. Our results set the north star for display development, with implications for future imaging, rendering and video coding technologies.

\end{abstract}

\keywords{visual resolution \and retinal resolution \and display resolution \and human visual acuity}

\begin{bibunit}[unsrtnat]

\section*{Introduction}

As display resolutions approach the limits of human perception, often known colloquially as the ``\emph{retinal resolution}'' or ``\emph{eye-limiting resolution}'', the law of diminishing returns begins to apply and further growth produces negligible improvements in perceived quality. The focus of our study is to measure this visual end-point of display resolution. In contrast, previous works on retinal or visual system resolution targeted individual mechanisms, for example, by eliminating optical aberrations \cite{thibos1987retinal}, or focused on visual system performance in motion \cite{anderson1991human}, shape \cite{weymouth1958visual} discrimination, or in Vernier acuity tasks \cite{levi1985vernier}. Our goal is to determine the resolution limit for a high-quality display, which appears indistinguishable from a perfect reference. Beyond the resolution limit of foveal vision, our work also sheds light on important ancillary variables unexplored in previous research, such as the resolution limits in retinal eccentricity (fovea versus periphery), and for (isoluminant) colour modulations.

The resolution limit of the eye is influenced by both optical and neural factors, which interact in a complex and often unintuitive manner. When the light enters the eye it is scattered in the ocular media\cite{navarro2009optical} and imperfectly focused on the retina due to optical aberrations\cite{yi2011depth, thibos1992chromatic}. The diffraction on the pupil restricts the maximum frequency that the eye can resolve\cite{charman1991wavefront, rovamo1998foveal}. Because of errors in accommodation, the resolution limit can vary with the viewing distance \cite{Depalma_Lowry_1962,Schober_Hilz_1965,Hernandez_Domenech_Segui_Illueca_1996}, resulting in a higher resolution limit at larger distances. We discuss this effect in the context of our study in \suppsecref{view-distance}. The resolution is also limited by the spacing of the photoreceptors\cite{hirsch1989spatial} and retinal ganglion cells\cite{kwon2019linkage}. Cones are densely packed in the fovea and decrease in density towards the periphery\cite{curcio1990human}. Retinal ganglion cells (RGCs), which pool information from the photoreceptors, show a similar distribution across the retina\cite{curcio1990topography}. Thus depending on the stimulus colour modulation and its position on the retina, any combination of optical and neural anatomical factors can be limiting the resolution of the eye\cite{campbell1966optical, snyder1986optical, williams1983consequences}. 

The key challenge of measuring the resolution limit is to provide precise control of the resolution (spatial frequency) of the displayed stimulus. A stimulus shown on an electronic display can be precisely reproduced only for the display's native resolution and its integer downsampling factors. Showing intermediate resolutions requires digital resampling, which alters the frequency content of the displayed stimulus. To avoid this problem, we employ a mechanised apparatus in which we can move the display towards or away from the observer, which lets us smoothly control the displayed resolution. By doing so, we reproduce the 130-year-old study of Wertheim (1894) \cite{wertheim1894ueber, wertheim1980peripheral} who used wire gratings to determine the relative drop of the resolution limit in parafoveal vision. Modern technology lets us employ robust psychophysical techniques, provide absolute measures of resolution, and include measurements for both achromatic and chromatic patterns.

The quantitative findings of this study have wide-ranging implications across multiple fields. For consumer electronics, understanding the limits of perceptible resolution can guide the development of more efficient and cost-effective displays. In virtual and augmented reality (VR/AR/XR), the optimisation of visual content based on perceptual thresholds, particularly in terms of retinal eccentricity and chromatic modulation, can enhance user experience without unnecessary computational costs. Additionally, in video compression, evidence-based chroma subsampling schemes can improve coding performance without affecting visual quality.

\section*{Results}

\begin{figure}[!b]
    \centering
    \hspace{-10mm}
    \includegraphics[width=\textwidth]{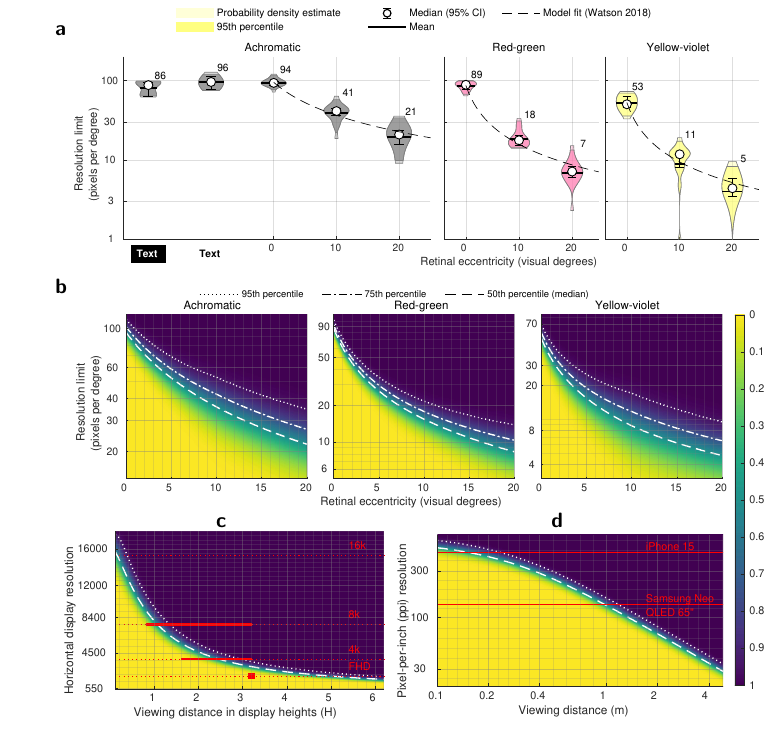}
    \caption{Spatial sensitivity and resolution limits for various colour directions across the visual field. \textbf{a.} The measured resolution limit in pixels-per-degree (ppd) at each eccentricity across the sample (N=18), with median (open circles), 95\% Confidence Intervals (CIs; error bars), and mean (horizontal bars). Numbers next to the violins indicate median ppd values of the observed data. Dashed lines represent the model fit. The edges of the shaded violin plot areas indicate the 95th percentile of thresholds. \textbf{b.} Heatmap showing the cumulative probability density of resolution limits within the observer sample, centred around predictions from the fitted Watson (2018)\cite{watson2018field} model. \textbf{c.} Ideal display vertical resolution as a function of viewing distance expressed in display height (H). The red horizontal bars indicate the ITU-R BT.2100-2\cite{bt2100} recommended viewing distances for various display resolutions: FHD (2K), 4K, and 8K. \textbf{d.} Required pixel-per-inch (ppi) resolution needed as a function of viewing distance (meters). In plots \textbf{b-d}, blue areas indicate that any further increase in pixel resolution would not be perceptible to almost all observers, while yellow areas represent resolutions that will be within the visual perceptual limit of almost all observers. The dotted and dashed lines represent different percentiles of the sample as shown in the legend.}
    \label{fig:min-ppd-ecc}
\end{figure}

\subsection*{Foveal resolution limit}

\figref{min-ppd-ecc}-a reports the measured resolution limit for achromatic, red-green, and yellow-violet patterns. The experiment measured the highest resolution that can be reliably detected by an observer at various eccentricities from the fovea. The resolution is expressed in pixels per degree or ppd. The corresponding maximum spatial frequency in cycles per visual degree is equal to half of the reported ppd values. 
The thresholds are reported for detecting a high-contrast Gabor patch as well as for identifying a decrease in resolution in black text on a white background, and white text on a black background simulating dark mode. The detailed breakdown of the mean, median, and 95th percentile threshold ppd values observed across different colour directions and eccentricity levels are provided in \tableref{mean_median_ci} of the \suppsecref{raw-data}. First, we will focus on the results for foveal vision, corresponding to an eccentricity of 0$^\circ$.

The widely accepted 20/20 vision standard, established by Snellen, suggests that the human eye can resolve detail at an angular resolution of 1 minute of arc, which corresponds to 60 pixels per degree (ppd) (\suppsecref{acuity-units} details the conversion between these units). This measure is derived from the design of the Snellen chart, where the smallest letters on the 20/20 line subtend an angle of 5 minutes of arc, with each critical feature of these letters subtending 1 minute of arc when viewed from 20 feet or 6 meters \cite{snellen1897methods, snyder1962herman}. This 1 arc minute value has historically been considered the threshold for human visual resolution, discussed in more detail in \suppsecref{historic-va}, leading to the assumption that 60\,ppd is sufficiently high for display purposes. However, younger observers with no optical abnormalities usually have acuities better than 20/20. In the context of displays, the Ultra Retina XDR display found in the 7th generation Apple iPad Pro (2024, 13") has an effective resolution of 65\,ppd when viewed from 35\,cm away, the shortest comfortable viewing distance. Both the 20/20 assumption and the Retinal display resolutions are significantly lower than the population mean of 94\,ppd we measured in our experiment, or individual values as high as 120\,ppd, as shown in \figref{min-ppd-raw} in \suppsecref{raw-data}. This demonstrates that the 60--65\,ppd range is not the \emph{retinal resolution} for a display. Note that high-contrast content, such as our Gabor patch stimulus, is not outside of the norm for content typically seen on displays: notably, text is often rendered at maximum contrast. To demonstrate, we also measured the detection threshold for text, both black-on-white and white-on-black (dark mode), and obtained values closely matching the resolution limit for sinusoidal gratings, as indicated by the "Text" data points in \figref{min-ppd-ecc}-a. Our results clearly indicate that the resolution limit of the eye is higher than broadly assumed in the industry.

It could also be surprising that the foveal resolution limit of red-green patterns is similar to that of achromatic patterns --- 89\,ppd for red-green vs. 94\,ppd for achromatic. It must be noted, however, that we did not try to isolate observers' individual chromatic mechanisms via the heterochromatic flicker paradigm \cite{Wagner1972}, as we wanted to capture data that could generalise across the population. Our results cast doubt on the common practice of chroma sub-sampling found in almost every lossy image and video format, from JPEG image coding to h265 or AV1 video encoding. The assumption of chroma subsampling is that the resolution of chromatic channels can be reduced twofold in relation to the achromatic channel due to the lower sensitivity of the visual system to high-frequency chromatic contrast. Our data suggests that this only holds for the yellow-violet colour direction, with the maximum resolution of 53\,ppd, but not for the red-green direction, consistent with the vision science theory that the isoluminant red-green pathway is the most sensitive opponent-colour channel of the human visual system\cite{chaparro1993colour}.

\subsection*{Resolution limit in periphery}

\figref{min-ppd-ecc}-a shows the rapid decline of the resolution limit as the stimulus was presented at increased eccentricities. This is in line with the established understanding that visual acuity and colour discrimination decrease as the stimulus moves away from the fovea, primarily due to the fall-off in cone density and the increase in receptive field size in the retina \cite{anstis1974chart, levi1985vernier}.  

The notable aspect of our results is that the resolution limit declines with increased eccentricity differently across colour directions. The achromatic resolution limit declines 2.3$\times$ between foveal vision and 10$^\circ$ eccentricity, while red-green declines 4.9$\times$ and yellow-violet 4.8$\times$. Popular techniques, such as foveated rendering \cite{Guenter_2012,Kaplanyan_2019} or foveated compression \cite{Illahi_2020}, are optimised for achromatic vision. Our results suggest that these techniques could provide further computational and bandwidth savings by lowering the resolution requirements for the chromatic channels. 

\subsection*{Modelling the resolution limit}

To interpolate and extrapolate our measurements, we fit the contrast sensitivity model presented by Watson (2018)\cite{watson2018field}:
\begin{equation}
\log(S^{\text{c}}(e, \rho)) = \log(\Sconscol{c}) + \krhocol{c}(1+\kecccol{c}e)\rho, \quad \forall \text{c}\in\{\text{Ach}, \text{RG},\text{YV}\},
\label{eq:res-limit-s}
\end{equation}
where $S^c$ is the contrast sensitivity of the colour channel $c$ for a given stimulus at eccentricity ($e$) and spatial frequency ($\rho$). $\Scons$ is the baseline sensitivity affected by other stimulus parameters (luminance, temporal frequency, size, etc). $\krho$ and $\kecc$ are the parameters of the model representing linear decrease with respect to spatial frequency and retinal eccentricity respectively. The contrast sensitivity $S$ is the inverse of the contrast of the stimulus. In our study, the contrast value is fixed for each colour direction (values reported in \tableref{stimuli-col} in \suppsecref{stimuli}). We optimise the values of $\Scons$, $\krho$ and $\kecc$ to predict the measured $\rho$ values from our data. The rearranged equation predicting the spatial frequency threshold as an inverse factor of eccentricity follows:
\begin{equation}
\rho(e) = \log{\left(\frac{S^{\text{c}}}{\Sconscol{c}}\right)}\frac{1}{\krhocol{c}(1+\kecccol{c}e)}, \quad \forall \text{c}\in\{\text{Ach}, \text{RG},\text{YV}\}.
\label{eq:res-limit-rho}
\end{equation}
The fitted model is drawn as dashed lines in \figref{min-ppd-ecc}-a. More details of the fitting procedure and the parameter values are provided in \suppsecref{model-fits}.

\subsection*{Resolution limit across the population}

In practical applications, it is important to know how the resolution limit varies across the population. This lets us make decisions that are relevant for the majority of the population. For example, designing a display which has \emph{retinal resolution} for 95\% of people rather than an average observer. To model the variation of the resolution limit in populations, we used the model from \eqref{res-limit-rho} to find the mean threshold, and then fitted a normal distribution to the per-observer data. To estimate the probability distribution at eccentricities not measured in our dataset, we linearly interpolated the parameters of the Gaussian distribution as detailed in \suppsecref{prob-dist-model}. The cumulative distribution, shown in \figref{min-ppd-ecc}-b, demonstrates a large variation across the population, especially at eccentricity. For example, if a median observer can see up to 22\,ppd at 20$^\circ$ eccentricity, this value increases to 35\,ppd for the 95th percentile of our sample. This shows the importance of considering individual differences in populations when designing technology aligned with human vision. Additionally, we also tested the effect of viewing distance on the resolution limit but did not observe a consistent trend among our sample of observers. More details of this investigation are discussed in \suppsecref{view-distance}.

We may also want to know how the resolution limit, expressed in ppd units, translates to actual displays and viewing distances. This is shown in \figref{min-ppd-ecc}-c, where we plot the relationship between the display resolution (number of horizontal lines) and the viewing distance (measured in display heights). Our model predictions can be compared with the ITU-R BT.2100-2\cite{bt2100} recommended viewing distances for television, shown as red horizontal lines in \figref{min-ppd-ecc}-c. Since Full HD (FHD) resolution was not designed to deliver a perfect image, the ITU recommendation of 3.2 display heights falls short of the reproduction below the visibility threshold. Our model indicates that a distance of at least 6 display heights would be necessary to satisfy the acuity limits of 95\% of the observers. For 4K and 8K displays, the ITU suggests viewing distances of 1.6--3.2 and 0.8--3.2 display heights, respectively. Our model shows that those ranges are overly conservative and there is little benefit of 8K resolution when sited further than 1.3 display heights from the screen. Used in this way, our model provides a framework to update existing guidelines and to establish new recommendations based on the limitations of our vision. In \figref{min-ppd-ecc}-d we plot the relation between pixel density (in pixels-per-inch) and viewing distance and show the screen resolution for two different devices.  To allow the readers to test their own displays, we created an online display resolution calculator available \href{https://www.cl.cam.ac.uk/research/rainbow/projects/display_calc/}{here}.


\begin{figure}[!b]
    \centering
    \begin{subfigure}[b]{\textwidth}
        \centering
        \includegraphics[width=\linewidth]{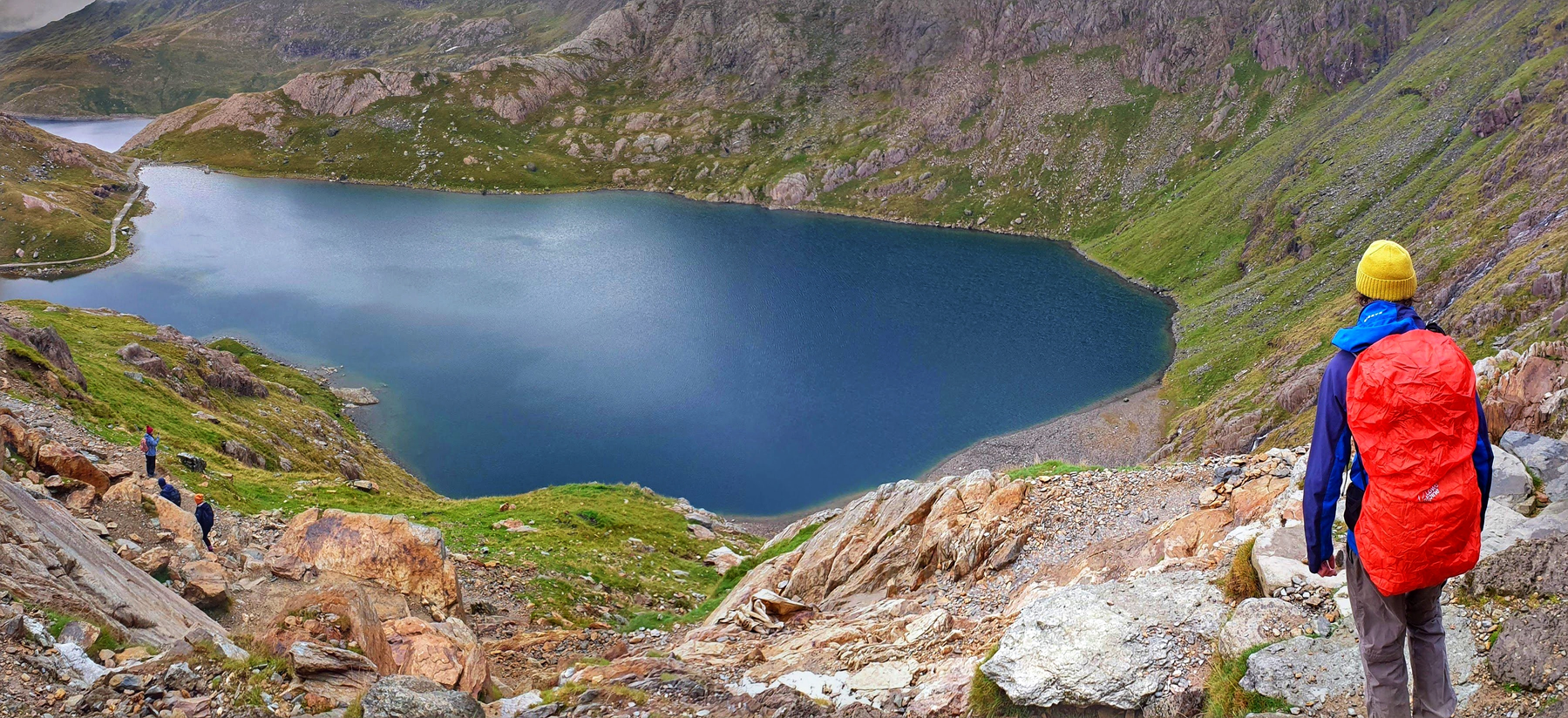} 
        \caption{Original image}
        \label{fig:img_original}
    \end{subfigure}
    \begin{subfigure}[b]{\textwidth}
        \centering
        \includegraphics[width=\linewidth]{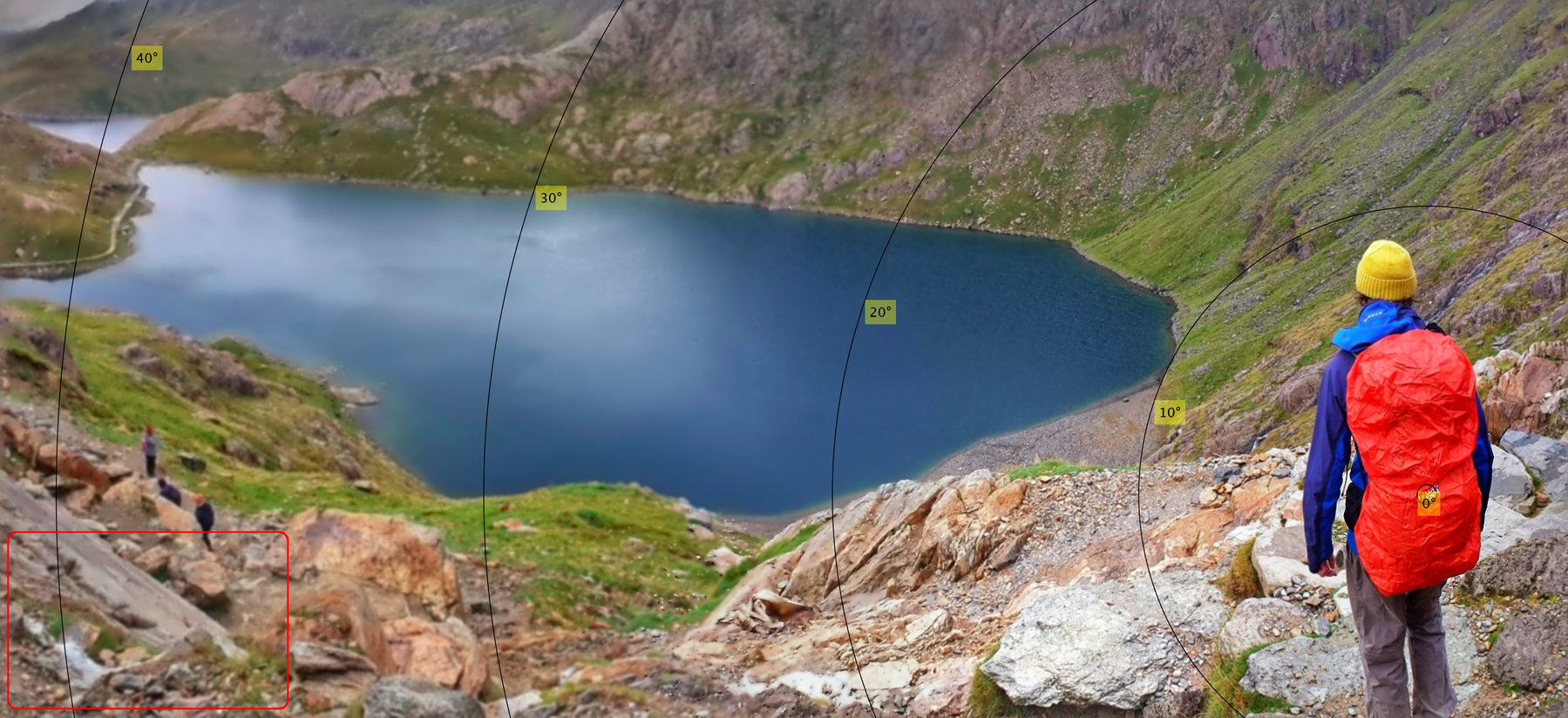} 
        \caption{Filtered image}
        \label{fig:img_simulated}
    \end{subfigure}
    \setlength{\abovecaptionskip}{-5pt}
    \caption{Eccentricity-dependent filtering that removes invisible details to improve coding performance. The contour lines show the retinal eccentricity positions relative to the gaze position. The filter was applied uniformly across discrete segments of eccentricity to better show the differences. Please zoom the page on the screen such that the red rectangle in the bottom-left corner of the simulated image is approximately the size of a credit card and view the image from 50\cm away. When the gaze is centred on the red backpack in the image, the degradation of high-frequency details in the periphery, will not be noticeable to the human eye.}
    \label{fig:graphics-appl}
\end{figure}

\subsection*{Example: Foveated rendering}
\label{sec:fov-render}

Foveated rendering, found in many commercial XR headsets, reduces the quality of rendered content depending on how far is a portion of the screen from the gaze location \cite{Guenter_2012,Kaplanyan_2019}. Foveated rendering typically reduces the resolution of rendered content to save bandwidth and the computational cost of rendering. The majority of foveated rendering methods consider only the perception of achromatic contrast and are manually tuned. Here, we show how our measurements could be used to find the right thresholds for foveated rendering for achromatic and chromatic contrast.

Here, we consider a simplified task of foveated filtering, which could improve video compression in foveated streaming -- we remove the high-frequency contrast that is invisible to the human eye. To produce the result shown in \figref{graphics-appl}, we decompose an image into achromatic, red-green and yellow-violet components of the DKL colour space \cite{Derrington1984}, decompose it into frequency bands using a Laplacian pyramid \cite{Burt_Adelson_1983}, and then set to zero the coefficients below the threshold contrast for a given eccentricity according to our fitted model for an average observer. More details of the image simulation can be found in \suppsecref{img-sim}. When the reconstructed image in \figref{graphics-appl} is seen from the right distance, and the gaze is directed towards the 0$^\circ$ target, the loss of resolution at larger eccentricities should be invisible. 


\begin{table}[!b]
\centering
\begin{tabular}{|>{\raggedright\arraybackslash}p{1.8cm}|p{1.8cm}|p{2cm}|>{\raggedright\arraybackslash}p{1.8cm}|>{\raggedright\arraybackslash}p{2.9cm}|>{\raggedright\arraybackslash}p{3.8cm}|}
\hline
\textbf{Study} & \textbf{Eccentricity range} & \textbf{Colour modulation} &  \textbf{Diameter in visual degrees } & \textbf{Data reported as} & \textbf{Stimuli \& methodology} \\
\hline
Wertheim (1894)\cite{wertheim1894ueber, wertheim1980peripheral} & 0\degree - 70\degree & Achromatic & 1\degree - 3\degree & Relative visual acuity (temporal) for 1 observer & Gratings made of high-precision wire frames\cite{strasburger2011peripheral} mounted on a moving rig viewed binocularly \\ \hline
Weymouth et al. (1928)\cite{weymouth1928visual} & 0\degree - 85' & Achromatic & 2.13\degree & Minimum angle of resolution (MAR), averaged across all meridians for 1 observer, reported in Weymouth (1958) \cite{weymouth1958visual} &  Square-wave interference fringe (Ives visual acuity object) with variable spatial frequency viewed monocularly \\ \hline
Ludvigh (1941)\cite{ludvigh1941extrafoveal} & 0\degree - 10\degree & Achromatic & - & Snellen fraction averaged for 3 observers & Individual Snellen test letters viewed monocularly  \\ \hline
Weymouth (1958)\cite{weymouth1958visual} & 0\degree - 20\degree & Achromatic & - & Minimum angle of resolution averaged for 20 observers &  Landolt C ring viewed monocularly \\ \hline
Kerr (1971) \cite{kerr1971visual} & 0\degree - 30\degree & Achromatic & 3\degree (square) & Log visual acuity values averaged for 2 observers & Square wave grating viewed monocularly via a Maxwellian apparatus \\ \hline
Anstis (1974) \cite{anstis1974chart} & 4\degree - 55\degree & Achromatic & - & Threshold letter heights in visual degrees averaged for two observers & Letters in a custom radial chart viewed binocularly \\ \hline
Thibos et al. (1987) \cite{thibos1987retinal} & 0\degree - 35\degree & Achromatic & 3\degree, 2.5\degree & Spatial resolution in cycles per degree (cpd) & Sinusoidal interference fringe directly on the retina viewed monocularly  \\ \hline
Anderson et al. (1991) \cite{anderson1991human} & 0\degree - 55\degree & Achromatic, red-green & 0.23\degree - 7.5\degree & Spatial resolution in cycles per degree (cpd) & Sinusoidal gratings viewed binocularly in motion discrimination and detection tasks \\ \hline
Masaoka et al. (2013) \cite{Masaoka2013} & Free viewing & - & 3.2\degree (square) & Spatial resolution in cycles per degree (cpd) & Colour images displayed at different resolutions compared with real objects \\ \hline
\end{tabular}
\setlength{\abovecaptionskip}{5pt}
\caption{Visual acuity measurements from literature. Please refer to \suppsecref{acuity-units} for conversion between different units of visual acuity.}
\setlength{\belowcaptionskip}{5pt}
\label{tab:literature}
\end{table}

\section*{Discussion}

As the aim of our experiments was to determine the ultimate resolution of a display, our methods and results differ from those found in studies on the resolution limit of the visual system. Here, we briefly discuss the most relevant studies, listed in \tableref{literature}. We focus on psychophysical measurements that consider all stages of vision but we do not review physiological measurements of individual mechanisms (e.g., retinal ganglion cell receptive field sizes \cite{harter1970evoked} or cone distribution and size in peripheral vision \cite{green1970regional}). 

Our methodology precisely controls the pixel-per-degree resolution incident on the visual system by changing the viewing distance, as well as resampling images by integer factors. This lets us explore the visual system's response to different spatial frequencies under controlled conditions. This approach is reminiscent of Wertheim's study from 1894 \cite{wertheim1894ueber, wertheim1980peripheral}, which highlighted differences in visual acuity across the visual field, but only reported the relative measurements of visual acuity. Weymouth et al. (1928) \cite{weymouth1928visual} reported carefully measured visual acuities with variable frequency gratings but only for the central 1.5\degree field of view. Others have utilised context-rich stimuli like Landolt rings \cite{weymouth1958visual}, letters \cite{ludvigh1941extrafoveal, anstis1974chart}, or images \cite{Masaoka2013}. Such measurements explain the resolution needed to achieve specific discrimination performance (e.g., determine the orientation of "c"), but these results do not directly translate into image quality.

\begin{figure}[!b]
    \centering
    \includegraphics[width=\columnwidth]{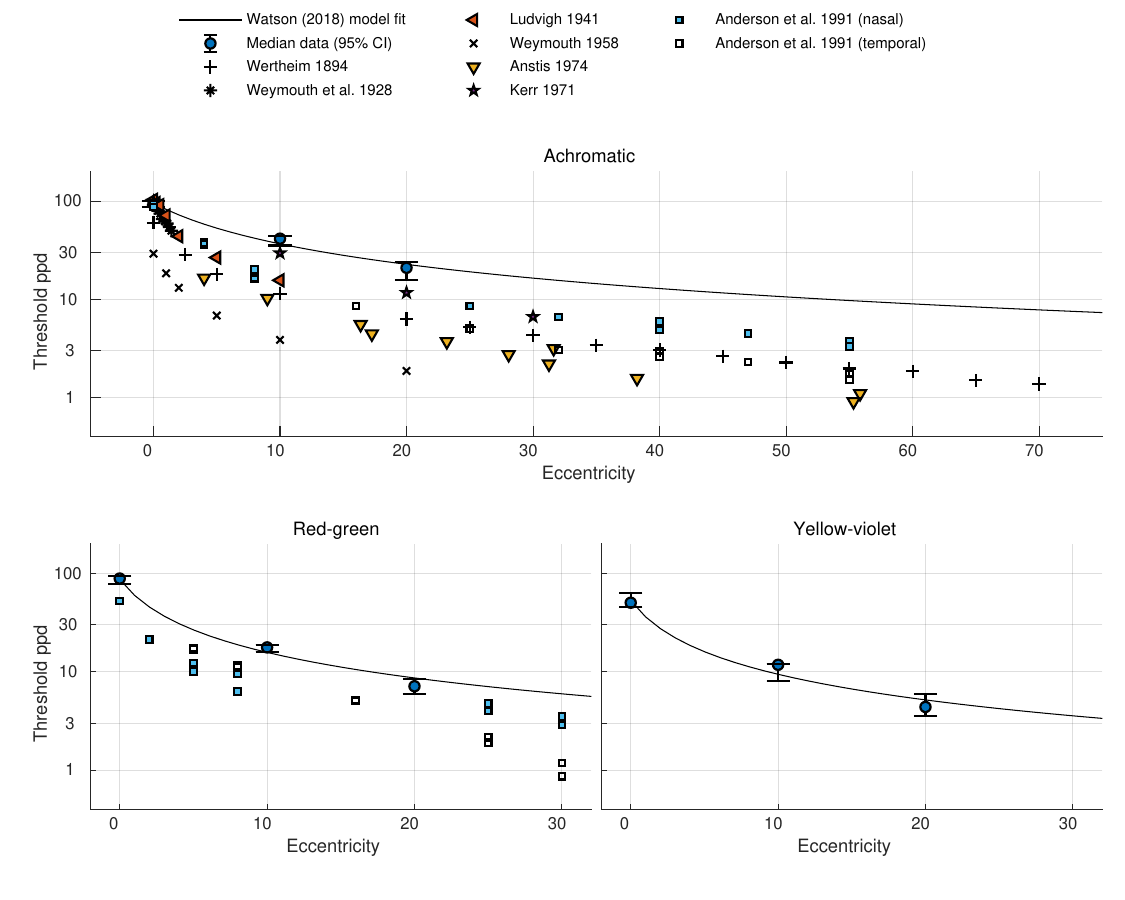}\\
    \setlength{\abovecaptionskip}{-12pt}
    \caption{Threshold pixel-per-degree (ppd) comparison with other models and datasets from the literature. }
    \label{fig:min-ppd-ecc-comp}
\end{figure}

Some studies attempted to isolate optical and neural factors and used specialised techniques such as interference fringes to project stimuli directly onto the retina, bypassing the eye's optics \cite{thibos1987retinal}, or used Maxwellian systems to converge light onto the pupil \cite{kerr1971visual}. These methodologies can help us understand which part of the visual system is the bottleneck, but they do not account for the everyday viewing experiences where the eye's optics play an important role. Our method addresses this by assessing the visual system's response as a whole, considering both the optics of the eye and the processing capabilities of the retina and higher neural mechanisms. A distinction should also be made between detection and discrimination measurements. For example, Anderson et al. (1991) \cite{anderson1991human} used a direction discrimination paradigm, and Healy \& Sawant (2012) \cite{healey2012limits} reported discrimination thresholds for different shapes, colour, and size features. Detection thresholds for simple stimuli, which only involve the observer perceiving the stimulus, are often higher than discrimination thresholds, where the observer must not only perceive but also identify and discriminate the stimulus (in terms of orientation, motion direction, etc.). Anderson et al. (1991) \cite{anderson1991human} also eliminated artefacts induced by high spatial frequency aliasing, so their measurements do not include stimuli that can still be detected but are perceived as distorted or aliased artefacts.

The resolution limits reported here capture the most conservative resolution limit --- the point at which even a high contrast stimulus could not be detected. \figref{min-ppd-ecc-comp} provides a direct comparison of resolution thresholds from datasets whose methodologies were comparable with ours. Most of these studies report lower resolution limits than those measured in our study, particularly in the peripheral regions. In contrast to our controlled threshold study, Masaoka et al. (2013) \cite{Masaoka2013} measured the ``realness'' scores for complex scenes at a fixed set of resolutions, which were produced by digital resampling. They observed that the scores began tapering off between the 53 and 78 cycles per degree (cpd) resolutions. Based on this, they estimated a resolution limit of approximately 60\cpd but did not directly measure any thresholds at this resolution. It should also be noted that natural images (such as those used in their study) rarely contain high contrast at high spatial frequencies (due to an expected 1/f power distribution)\cite{burton1987color}, though artificial content, such as text, often does contain such high contrasts.

When comparing our results with historical data, it is evident that while our methodology leverages modern technological advances, the fundamental characteristics of human visual acuity documented in earlier works remain consistent. Our study reaffirms the findings of earlier investigations into retinal resolution but also expands the scope by employing a holistic approach that considers the entire visual system. This approach provides a more accurate representation of visual performance in real-world viewing situations, offering valuable insights for the design of displays and other applications where visual clarity and resolution are critical. 

\section*{Methods}

Below we summarise the main points of the methods, while the detailed description can be found in the supplementary. 

\subsection*{Moving display apparatus}

\begin{figure}
    \centering
    \hspace{-10mm}
    \includegraphics[width=\textwidth]{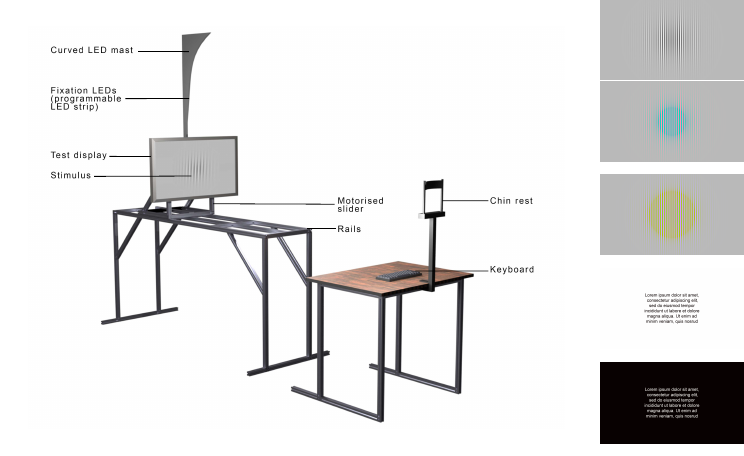}
    \caption{Left: Rendition of the experimental setup. The display can slide on rails towards and away from the observer. The movement is controlled by a motorised camera slider to present stimuli at different pixel-per-degree (ppd) resolutions. The fixation point for the foveal presentation is the black cross in the centre of the screen. For peripheral viewing, an LED on the curved LED mast is lit up corresponding to the retinal eccentricity. The curvature was designed to approximate the distance to the horopter (for the average display position). The photograph of the actual apparatus can be found in the supplementary information \figref{movable-display}. Right: Stimuli used in the experiments. From top-to-bottom: achromatic, red-green and yellow-violet square-wave gratings, black-on-white text, and white-on-black text. }
     \label{fig:movable-display-figure}
\end{figure}

We designed an experimental apparatus capable of adjusting the display's position relative to the observer to smoothly vary the effective resolution (in terms of pixels per visual degree) without the need for pixel resampling. The apparatus features a 27-inch Eizo ColorEdge (CS2740) 4K monitor mounted on a mobile cage, allowing it to move smoothly along a 1.6-meter track as shown in \figref{movable-display-figure}. This setup enables a wide range of viewing distances. The movement is controlled by a stepper motor via an Arduino Uno microcontroller. Central fixation for foveal stimuli is provided by a black cross at the centre of the display. For stimuli at larger eccentricities, the fixation point is adjusted using specific LEDs on a programmable LED strip controlled by a Raspberry Pi Pico microcontroller. 

\subsection*{Stimuli}

For the gratings experiments, the stimuli consisted of square-wave gratings, modulated with a Gaussian envelope, with an underlying frequency equal to the Nyquist frequency of the display. These stimuli are widely used in vision science experiments because the foundational visual detectors of the human visual system are likely optimised for similar waveforms\cite{watson1983does} The gratings were modulated along three directions in the DKL colour space \cite{Derrington1984}: achromatic (L+M) with a contrast of 0.96, chromatic (L-M) with a contrast of 0.23, and chromatic (S-(L+M)) with a contrast of 0.89. The contrasts were chosen to ensure they could be faithfully reproduced on the display. The resolution (spatial frequency) of the stimuli was modulated by moving the display towards or away from the observer and by upsampling or downsampling the spatial frequency of the Gabor patch using integer factors of 2, 3, and 4. The stimuli were shown at three eccentricities: 0\degree{}, 1\degree{} and 20\degree{}, and at luminance level 100\cdms{}.

A different procedure was used for the experiment involving text stimuli. We created two images with the same text but in opposite polarities; one with a white background and black text, and the other with a black background and white text. The contrast between the text and the background in both images was 0.96. To ensure that the text appeared the same size regardless of the viewing distance, a high-resolution reference image was resampled to the native resolution of the display using a Lancosz filter ($a=3$). The images spanned 12.63\degree${\times}$7.12\degree visual degrees at all viewing distances. The test stimulus was additionally resampled to match the required pixel resolution for the trial using the corresponding $n{\times}n$ box filter. This nearest-neighbour interpolation effectively simulates a lower (than display native) pixel resolution. For the text experiment, the luminance of the white background was 191\cdms and for the black background, it was 8.67\cdms. Only foveal stimuli with free viewing were used for the text stimuli. The stimuli used in our experiments are shown in \figref{movable-display-figure}. More details of the stimuli are provided in \suppsecref{stimuli}.

\subsection*{Psychophysical Method}

We used a two-interval-forced-choice (2IFC) paradigm. Participants were seated in a dark room, facing the display such that the viewing distance was between 110 and 270\,cm depending on the values of the tested \ppd for the stimuli. Each trial began with a masker presentation, followed by two intervals, one containing the stimulus and the other blank. Participants indicated which interval contained the stimulus, and their responses were recorded. Each of the trials was repeated 3 times consecutively to reduce false positive and false negative responses. The QUEST\cite{watson1983quest} adaptive procedure was used to select the next pixel-per-degree resolution to be tested based on participants' responses. The threshold for each observer was estimated from  30 to 50 QUEST trials. The trials stopped when either the maximum allowed number of trials was reached or the standard deviation of the threshold estimate determined by \code{QuestSd} function from Psychtoolbox\cite{kleiner2007s} reached a value of 0.07. The value of the stopping criterion was chosen to ensure sufficient precision in the threshold estimate while keeping the experiment duration manageable and was determined based on initial testing with a small group of observers.

\subsection*{Observers}

Eighteen observers (6 female, 12 male) with a mean age of 25.5 (age range: 13-46 years) participated in the first part of the experiment (gratings stimuli). For the second part of the study (text stimuli), twelve of the original participants (3 female, 9 male) were involved. All had normal or corrected-to-normal vision and normal colour vision, tested with Ishihara colour plates. Visual acuities were tested using Snellen charts (more details in \suppsecref{snellen-chart}). The experiment was approved by the departmental ethical committee at the Department of Computer Science and Technology of the University of Cambridge. Participants were recruited through university mailing lists, and most were students or members of the lab group. Before participation, all observers were briefed on the purpose of the research, the procedures involved and their rights as participants, including the right to withdraw at any time without penalty. Each participant read and signed a standard consent form outlining these rights before the experiment. The experiment was also demonstrated to the participants and they were asked to do a short trial run to ensure they understood the task before the main experiment started. All the data was anonymised and the participants were compensated for their time.

\subsection*{Data Analysis}
\label{sec:data-analysis}

We used the Maximum Likelihood Estimation (MLE) method to fit psychometric functions to individual participants' binary responses from the QUEST trials. The likelihood function quantified how well a psychometric function explained the observed responses across different stimulus intensities. Outliers, identified based on the modified Z-score, were excluded from the dataset to prevent skewed estimates of psychophysical trends. In total, 13 out of 162 data points were identified as outliers and excluded from subsequent analyses. The post-hoc analysis showed that those were most likely caused by lapses of attention. The individual data points, including the outliers, are shown in \figref{min-ppd-raw} in the \suppsecref{raw-data}.

\addtocontents{toc}{\protect\setcounter{tocdepth}{-1}}


\putbib[references]

\section*{Acknowledgements}

We would like to extend our gratitude to Thomas Bytheway for his exceptional work in building the display rig. We also thank Scott Daly for his feedback and suggestions.

\section*{Author contributions statement}

A.C. proposed the project and R.K.M. and A.C. conceived the experiment. M.A. designed the psychophysical study and conducted the experiment under R.K.M.'s supervision. M.A. analysed the results and created the figures with feedback from all authors. All authors contributed to writing and reviewing the manuscript.

\end{bibunit}

\newcounter{savepage}
\setcounter{savepage}{\value{page}}

\clearpage
\pagestyle{empty} 

\begin{bibunit}[unsrtnat]


\setcounter{page}{1}
\pagestyle{fancy} 
\fancyhf{} 
\fancyfoot[C]{\thepage} 

\noindent{\huge\sffamily\bfseries Supplementary Information}\\[1em] 

\addtocontents{toc}{\protect\setcounter{tocdepth}{2}}

\tableofcontents

\renewcommand{\thefigure}{\Alph{figure}}
\renewcommand{\thetable}{\Alph{table}}
\setcounter{figure}{0}
\setcounter{table}{0}
\setcounter{equation}{0}

\section{Historical context of visual acuity standards}
\label{supp:historic-va}

As explained by Velasco e Cruz (1990) \cite{e1990historical}, Snellen's widely used visual acuity chart was based on Helmholtz's (1962) \cite{helmholtz1962} report of the resolving power of the human eye, which was 1 minute of arc. However, this value actually represented twice the width of the wire grating used in Helmholtz's experiments, corresponding to a full square grating cycle. For the Snellen chart design, approximately 0.5 minutes of arc should have been used. The Snellen chart is designed such that the letters on the 20/20 line subtend an angle of 5 arc minutes, with each distinguishing feature subtending 1 arc minute of visual angle when viewed at 20 feet or 6 meters \cite{snellen1897methods, snyder1962herman}. This has led to the historical misconception that 1 arc minute, or 60 ppd, is the critical resolution of the human eye and is often wrongly considered sufficient for display purposes.

\section{Visual acuity units conversion}
\label{supp:acuity-units}

Visual acuity of angular resolution of the human eye is represented in many different units in the literature. Understanding these various units and how to convert between them is necessary for interpreting results and comparing data across different studies and contexts. The primary units of visual acuity include Snellen fraction, pixels per degree (ppd), and logMAR. Below are the mathematical formulae for conversion between these units, along with explanations.

\subsection{Snellen Fraction}

The Snellen fraction is one of the most common ways to express visual acuity. It is represented as a fraction, where the numerator indicates the testing distance (usually 20 feet or 6 meters), and the denominator indicates the distance at which a person with normal vision can read the same line of the eye chart. In simple terms, it is a ratio representing the visual acuity of an individual compared to normal vision.
\begin{equation}
\snellfrac = \frac{\dchart}{\dnorm},
\end{equation}
where, \snellfrac\, is the Snellen fraction, \dchart\, is the distance between the observer and the chart, and \dnorm\, is the distance at which a normal eye can read the line. For example, a Snellen fraction of 20/40 means that the test subject can read at 20 feet what a person with normal vision can read at 40 feet.

\subsection{logMAR}
LogMAR (logarithm of the Minimum Angle of Resolution) is a measure of visual acuity that provides a more intuitive scale. The logarithmic scale provides a more linear representation of visual acuity and is widely used in research and clinical settings. It is the minimum angle in minutes that can be resolved by an observer. The value is related to Snellen fraction as follows: 
\begin{equation}
    \log{\text{MAR}} = {\log}_{10}\frac{1}{\snellfrac}.
\end{equation}
For example, if the Snellen fraction is 20/40, then:
\begin{equation}
    \log{\text{MAR}} = {\log}_{10}\frac{40}{20} = {\log}_{10}(2) = 0.3010.
\end{equation}

To convert from logMAR to Snellen fraction:
\begin{equation}
\snellfrac = \frac{1}{10^{\log_{10}{\text{MAR}}}}.
\label{eq:snellen_to_logmar}
\end{equation}

\subsection{Pixels-per-degree}

Pixels per degree (ppd) is a measure of spatial resolution used in digital displays, indicating how many pixels occupy one degree of visual angle. The ppd can vary across the display due to differences in viewing angle at different points on the screen, particularly at the edges compared to the centre \cite{Mantiuk_2021_FovVideoVDP}. However, for practical purposes and consistency across studies, we use the ppd formula for the centre of the screen:
\begin{equation}
     \text{ppd} \approx \frac{\pi}{360\cdot\arctan{\frac{0.5\dwidth}{\rw \dv}}},
\end{equation}
where $\dwidth$ is the width of the screen, $\rw$ is the display pixel resolution along the width, and $\dv$ is the viewing distance between the observer and the display.
In our experiments, the stimulus was always located at the centre of the screen, which was far enough for the approximation to hold. The eccentricity measurements used the distance to the horopter, which ensured correct calculations.

In the context of resolving power, the pixel-per-degree value relates to the viewer's visual acuity limit. For example, 20/20 vision is equivalent to a minimum resolving angle of 1 arc minute. For displays, this would mean that a pixel, or the smallest resolvable element, should span 1/60th of a visual degree or 60\ppd to match this resolution limit. Conversely, 20/10 vision would correspond to 120\ppd, and so on. This relationship can be expressed as:
\begin{equation}
    \text{ppd} = 60 \cdot \snellfrac. 
\end{equation}
Similarly, the mathematical relationship between logMAR and ppd can be expressed by substituting from \eqref{snellen_to_logmar}:
\begin{equation}
    \text{ppd} = \frac{60}{10^{\log{\text{MAR}}}}
\end{equation}

\section{Methods}

Further details of the methodology that were not covered in detail in the main text are provided here, including additional figures. 

\subsection{Apparatus}

\begin{figure}[!htbp]
    \centering
    \includegraphics[width=0.8\columnwidth]{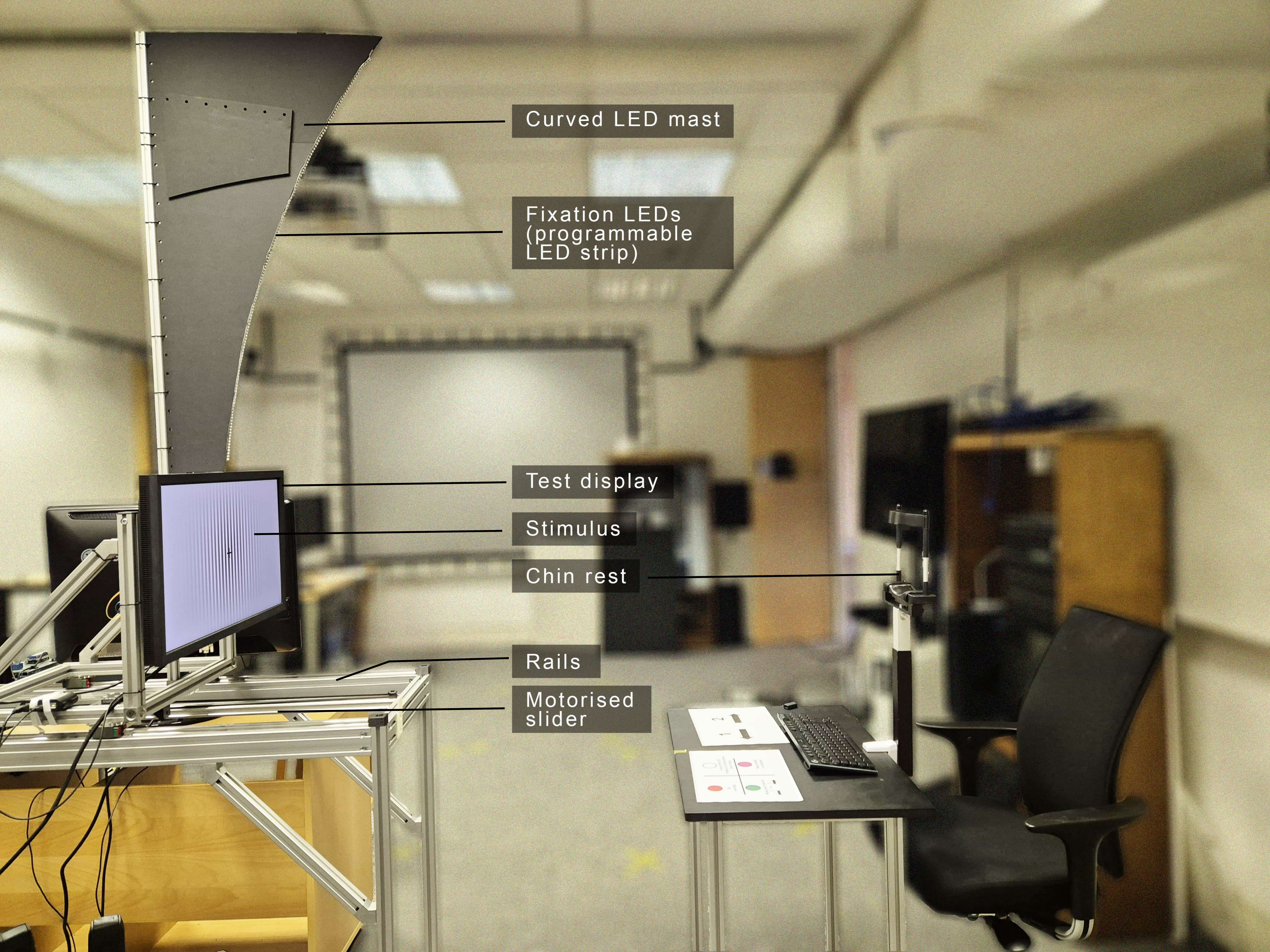}\\
    \caption{Experimental setup. The display can slide on the rails towards and away from the observer. The movement is controlled by a motorized camera slider to show stimulus at different pixel-per-degree (ppd) resolutions. The fixation point for the foveal presentation is the black cross in the centre of the screen. For peripheral viewing, an LED on the curved LED mast is lit up for the corresponding retinal eccentricity. }
    \label{fig:movable-display}
\end{figure}

\begin{figure}[!htbp]
    \centering
    \includegraphics[width=0.5\columnwidth]{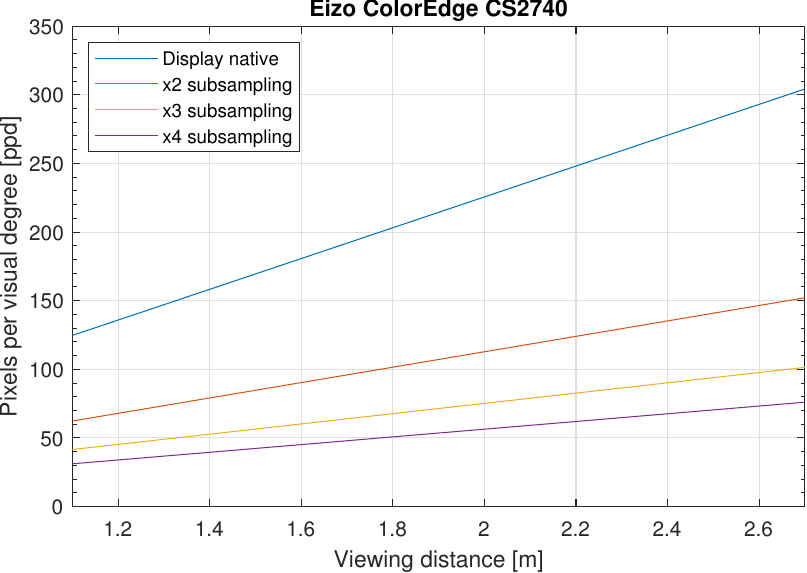}
    \caption{The resolution that Eizo ColorEdge CS2740 displays can reproduce at a range of viewing distances.}
    \label{fig:disp-ppd-coloredge}
\end{figure}

We used a moving display apparatus as described in Section \emph{Methods - Moving display apparatus} in the main text. \figref{movable-display} shows the photograph of the actual setup. \figref{disp-ppd-coloredge} illustrates the relationship between viewing distance and the effective resolution (ppd) that can be achieved with the moving apparatus. The graph shows how the display's native resolution changes as a function of viewing distance and the subsampled resolutions (×2, ×3, and ×4 subsampling). The advantage of this setup is that we are able to simulate an effectively continuous range of pixel-per-degree (ppd) values for our experiment. We can also simulate the same ppd values at different viewing distances by changing the subsampling resolution. This setup allows for a flexible and precise adjustment of the display’s resolution.

\FloatBarrier

\subsection{Stimuli}
\label{supp:stimuli}

\begin{figure}[!b]
    \centering
    \includegraphics[width=\textwidth]{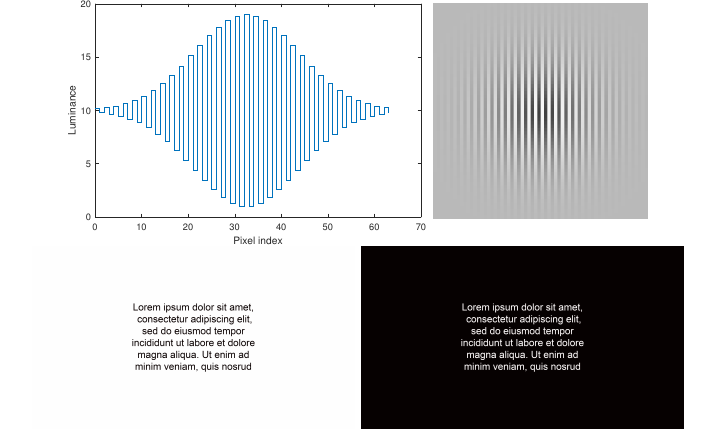} 
    \caption{a. A cross-section of the square-wave grating, modulated by a Gaussian envelope, generated at the Nyquist frequency. b. the image of the stimulus. c, d. Stimuli for text experiment.}
    \label{fig:gabor}
\end{figure}

As described in Section \textit{Methods - Stimuli}, we used square-wave gratings and text stimuli for our experiments. \figref{gabor} shows a cross-section of an achromatic grating stimulus and the text images. \tableref{stimuli-col} details the colour coordinates of the gratings used in our experiments. The units of contrast sensitivity and the cone contrast are the same as the ones used in Ashraf et al. (2024) \cite{ashraf2024castlecsf}.

\begin{table}[!b]
\centering
\caption{Stimuli colour information}
\setlength{\belowcaptionskip}{5pt}
\label{tab:stimuli-col}
\begin{tabular}{cccccc}
\toprule
\textbf{Contrast sensitivity} & \textbf{Cone contrast} & \textbf{Colour} & \textbf{Luminance (\cdms)} & \textbf{x} & \textbf{y}\\
\midrule
\addlinespace
\multicolumn{6}{c}{Achromatic} \\ 
\cline{3-3}
\multirow{2}{*}{1.09} & \multirow{2}{*}{0.913} & White & 191 & 0.3127 & 0.329 \\ 
  &   & Black  & 8.67 & 0.3127 & 0.329 \\ 
\midrule
\addlinespace
\multicolumn{6}{c}{Red-green} \\ 
\cline{3-3}
\multirow{2}{*}{7.42} & \multirow{2}{*}{0.135} & Red & 100 & 0.4022 & 0.2834 \\ 
  &   & Green & 100 & 0.2410 & 0.3710 \\ 
\midrule
\addlinespace
\multicolumn{6}{c}{Yellow-violet} \\ 
\cline{3-3}
\multirow{2}{*}{2.05} & \multirow{2}{*}{0.487} & Violet & 100 & 0.2756 & 0.2394 \\ 
  &   & Yellow & 100 & 0.3901 & 0.5157 \\ 
\bottomrule
\end{tabular}
\end{table}

The stimuli were shown at different pixel-per-degree values and the observers' response to whether they could detect the stimulus at the corresponding ppd values were recorded. To modulate the resolution of the stimuli we used the following two strategies:
\begin{itemize}
    \item Move the display towards or away from the observer using the mechanised camera slider (see \figref{movable-display});
    \item Upsample or downsample the spatial resolution of the screen using integer factors (2$\times$, 3$\times$, 4$\times$,...). The corresponding spatial resolutions are shown as 2$\times$, 3$\times$, and 4$\times$ subsampling lines in \figref{disp-ppd-coloredge}.
\end{itemize}
For each trial, we calculated the viewing distance and the resolution sub-sampling factor that would require the least amount of movement with respect to the display's current position. For example, if the display is currently at a viewing distance of 140\cm and we want to display a stimulus at 50\ppd for our next trial, some of our options within the range of motion of the display are to move the display: i) 51\cm towards the observer at 2x subsampling, ii) 7\cm towards the observer at 3x subsampling, and iii) 38\cm away from the observer at 4x subsampling. The second option in this case requires the display to move only 7\cm and would thus reduce the time required to physically move the display between consecutive trials.

\subsection{Prior psychometric function estimation}

\begin{figure}[!b]
    \centering
    \includegraphics[width=0.85\columnwidth]{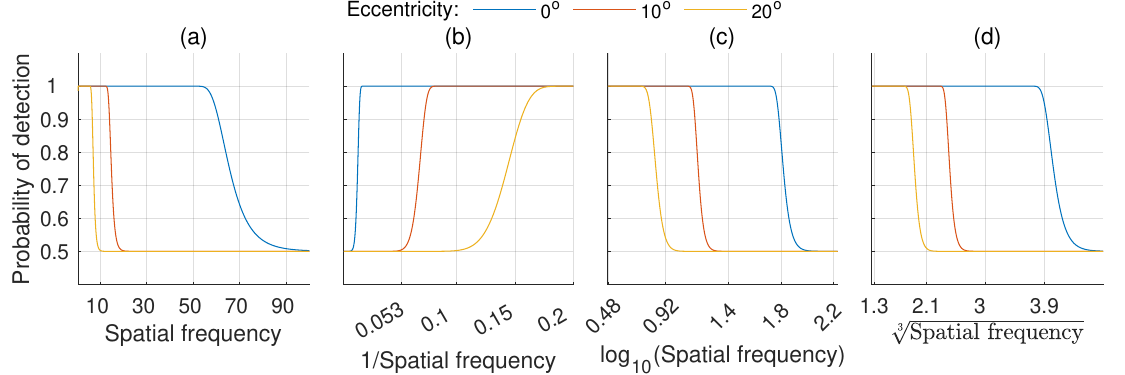}\\
    \caption{Psychometric function used to estimate spatial frequency detection thresholds across different eccentricities (0\degree, 10\degree, and 20\degree). The curves show the probability of detection as functions of spatial frequency, plotted using various transformations.}
    \label{fig:psych-func}
\end{figure}

In order to use an adaptive procedure (QUEST\cite{watson1983quest}), it is necessary to find a psychometric function whose shape remains the same regardless of the tested ppd value. We used a contrast sensitivity function (CSF) model, castleCSF \cite{ashraf2024castlecsf}, to estimate the likely psychometric functions for spatial frequency detection thresholds in a 2IFC experiment. The psychometric function used in the CSF model is a Weibull function w.r.t cone contrast value of the stimulus. For different spatial frequencies, the function is displaced along the contrast axis, but the slope of the function does not change. We generated a matrix of these psychometric functions for a dense range of spatial frequencies. Slicing this matrix for our required stimulus contrast effectively produces the psychometric functions w.r.t spatial frequency as shown in \figref{psych-func}(a). The main difficulty was to find the suitable function of spatial frequency for which the shape of the psychometric function remains consistent for different conditions. Initial attempts to represent spatial frequency on linear, logarithmic, and inverse scales (\figref{psych-func}(a-c) respectively) revealed varying psychometric function slopes, indicating that the typical assumption of constant slope for psychometric functions in contrast sensitivity experiments may not hold in experiments where spatial frequency is the parameter being measured. To find the optimum scaling, we ran an optimisation routine to minimise the differences in slopes with the exponent of the spatial frequency transformation as a fitted parameter. The value of this fitted parameter was close to 0.3, therefore, a transformation involving the cube root of spatial frequency was explored, which, based on the collected data, provided a more consistent psychometric function across different eccentricities and colour directions. We also used this transformation for scaling the threshold ppd values when fitting models and normalised probability distribution functions:
\begin{equation}
f(\rho) = \sqrt[3]\rho,
\label{eq:ppd-to-fppd}
\end{equation}
where $\rho$ is the spatial frequency in \cpd. The ppd values are twice the \cpd values. The resulting psychometric function can be seen in \figref{psych-func}(d).

\subsection{Psychophysical method}

The study employed a 2-Interval Forced Choice (2IFC) paradigm to measure the minimum resolution required to detect the stimulus. This psychophysical method is advantageous for its sensitivity and accuracy in measuring perceptual thresholds, as it minimises bias by forcing a choice between two presented intervals, one containing the test stimulus described in the section \textit{Stimuli} and the other a uniform field. Participants were seated comfortably in a dark room, facing the display screen at an initial distance of 190\,cm. Each trial began with the presentation of a masker for 400\,ms, followed by two intervals. The order of the stimulus and uniform field intervals were randomised across trials to prevent anticipation. Participants were instructed to indicate which interval contained the stimulus and their response was recorded as binary data, representing correct or incorrect decisions. The observers received feedback immediately after their choice.  

Each stimulus was presented for 500\,ms. The stimulus's onset and exit were modulated by a Gaussian function with $\sigma=200$\,ms. This was to ensure that higher temporal frequencies associated with the onset and exit did not facilitate the detection/discrimination. A sound was played at the onset of each stimulus. QUEST adaptive procedure was used to select the next pixel per degree resolution to be tested, based on the participant's responses. We collected data from 30 to 50 QUEST trials for each condition and each observer. During the display movement, a random noise pattern was shown on the display. The main experiment was completed in three blocks, one for each of the retinal positions that we tested: i) foveal (0\degree), ii) parafoveal 10\degree, and iii) parafoveal 20\degree. The order of the tested colour directions within a session was randomised. 

\subsection{Data analysis}

The individual observers' thresholds are provided in \figref{min-ppd-raw}. The outliers are the individual observations that deviate significantly from the group data, because of individual errors. These deviations can occur due to various factors, including individual variations in visual acuity, attentional lapses, or misunderstanding of task instructions. Particularly in peripheral viewing, where acuity drops markedly, small deviations in gaze direction—a common occurrence given the challenge of maintaining a steady eye position—can result in data points that lie far outside the typical response range. To ensure the robustness of our findings, these outliers were systematically excluded to prevent skewed estimates of the overall psychophysical trends. For each combination of colour direction and eccentricity condition, the median absolute deviation (MAD)\cite{leys2013detecting} of the data from all individual observers was computed. The modified Z-score was then assigned to each data point based on how many MADs away from the median it was. The data points with a modified Z-score greater than 3.5 were considered outliers and systematically removed from the dataset. In total, 13 out of 162 data points were identified as outliers and were excluded from subsequent analyses. 

\section{Results and discussion}

In this section, we provide additional details and data that complement the findings discussed in the main results section of our study. The supplementary materials include extended data tables and figures that show more details about the variability observed across different participants and conditions and more details on the parameters of the model.

\subsection{Individual observer data}
\label{supp:raw-data}

\begin{figure}[!t]
    \centering
    \includegraphics[width=0.85\columnwidth]{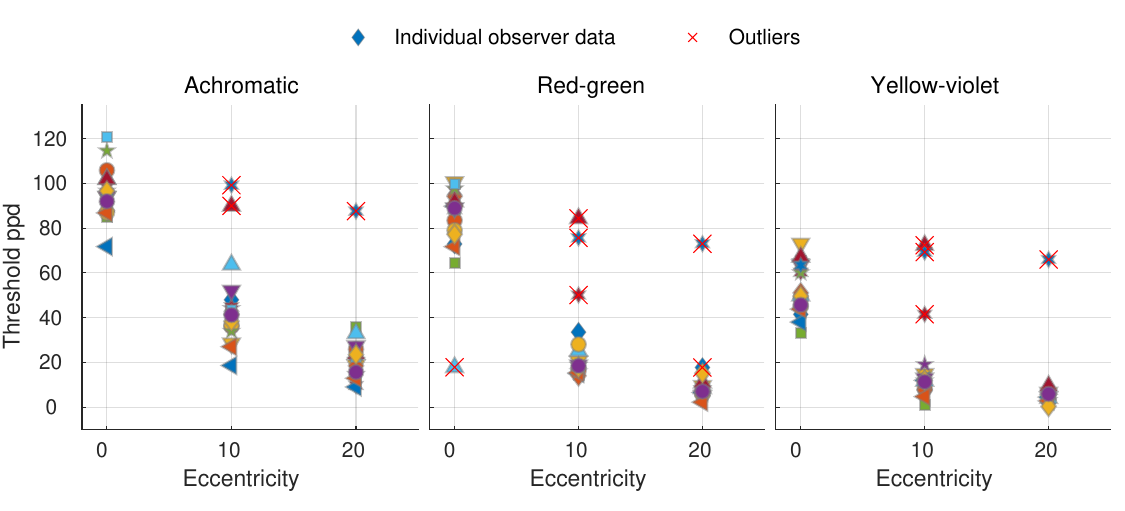}\\
    \setlength{\abovecaptionskip}{-2pt}
    \caption{Threshold ppd (pixels per degree) values from individual observers across varying retinal eccentricity (in visual degrees) for three colour channels: achromatic, red-green, and yellow-violet. Outliers, indicated by red crosses, were identified and excluded from subsequent analysis. Each colour channel demonstrates a trend of decreasing threshold with increasing eccentricity, reflecting a decline in spatial sensitivity away from the foveal centre.}
    \label{fig:min-ppd-raw}
\end{figure}

The raw 2IFC results were fitted with a maximum likelihood function (as described in the \emph{Methods - Data Analysis} section in the main text) for each observer to estimate their individual spatial frequency threshold. \figref{min-ppd-raw} shows the threshold ppd values from individual observers across varying retinal eccentricities for three colour channels: achromatic, red-green, and yellow-violet. Outliers, identified by red crosses, were excluded from subsequent analysis to ensure the robustness of the results. \tableref{mean_median_ci} shows the corresponding central tendency and variability of spatial frequency thresholds in pixels per degree (ppd) for the three colour channels at three levels of visual eccentricity (0\degree, 10\degree, 20\degree). The 'Mean' and 'Median' values across all observers, along with their respective 95\% Confidence Intervals (CIs), are reported. Additionally, the '95th percentile' values provide the range of threshold values for 95\% of the observers.  In both the individual and aggregated data, each colour channel demonstrates a trend of decreasing threshold with increasing eccentricity, reflecting a decline in spatial sensitivity away from the foveal centre.  This decline reflects the impact of both optical and neural factors on spatial resolution across different regions of the retina.

\begin{table}[!t]
  \centering
  \setlength{\abovecaptionskip}{5pt}
  \caption{The central tendency and variability of spatial frequency thresholds, in pixels per degree (ppd), for different colour channels (achromatic, red-green, yellow-violet) at three levels of visual eccentricity (0, 10, 20 visual degrees). The 'Mean' and 'Median' values across all observers along with the respective 95\% Confidence Intervals (CIs) are reported. The '95th percentile' values provide the range of threshold values for 95\% of the observers. The data shows a decrease in threshold with increasing eccentricity for each colour direction, indicating a consistent decline in visual sensitivity as the distance from the foveal centre increases.}
    \begin{tabular}{|l|l|llll|llll|rlr|}
    \hline
    \multicolumn{1}{|c|}{\textbf{Color direction}} & \multicolumn{1}{c|}{\textbf{Eccentricity}} & \textbf{Mean} & \multicolumn{3}{c|}{\textbf{[95\% CI]}} & \multicolumn{1}{c}{\textbf{Median}} & \multicolumn{3}{c|}{\textbf{[95\% CI]}} & \multicolumn{3}{c|}{\textbf{[95th percentile]}} \\
    \hline
    \multirow{3}{*}{Achromatic} & 0     & 95.37 & 90.45 & ,     & 100.7 & 94.01 & 87.78 & ,     & 100.2 & 76.65 & ,     & 117.2 \\
\cline{2-13}          & 10    & 40.15 & 35.19 & ,     & 45.21 & 41.40 & 35.33 & ,     & 44.39 & 22.04 & ,     & 58.93 \\
\cline{2-13}          & 20    & 20.81 & 17.28 & ,     & 24.51 & 20.93 & 15.69 & ,     & 24.18 & 9.796 & ,     & 34.06 \\
    \hline
    \multirow{3}{*}{Red-green} & 0     & 86.77 & 81.72 & ,     & 91.44 & 89.00 & 79.05 & ,     & 94.21 & 67.30 & ,     & 99.97 \\
\cline{2-13}          & 10    & 19.24 & 16.82 & ,     & 22.09 & 17.74 & 15.95 & ,     & 18.82 & 14.32 & ,     & 31.12 \\
\cline{2-13}          & 20    & 7.461 & 6.157 & ,     & 8.924 & 7.146 & 5.966 & ,     & 8.433 & 3.298 & ,     & 13.51 \\
    \hline
    \multirow{3}{*}{Yellow-violet} & 0     & 53.56 & 48.45 & ,     & 58.62 & 50.49 & 45.36 & ,     & 63.40 & 35.38 & ,     & 70.80 \\
\cline{2-13}          & 10    & 10.55 & 8.249 & ,     & 12.64 & 11.80 & 8.038 & ,     & 12.05 & 2.643 & ,     & 17.40 \\
\cline{2-13}          & 20    & 4.715 & 3.660 & ,     & 5.800 & 4.399 & 3.530 & ,     & 5.946 & 1.109 & ,     & 8.703 \\
    \hline
    \end{tabular}%
  \label{tab:mean_median_ci}%
\end{table}%


\subsection{Relationship with visual acuity}
\label{supp:snellen-chart}

\begin{figure}[!t]
    \centering
    \includegraphics[width=\linewidth]{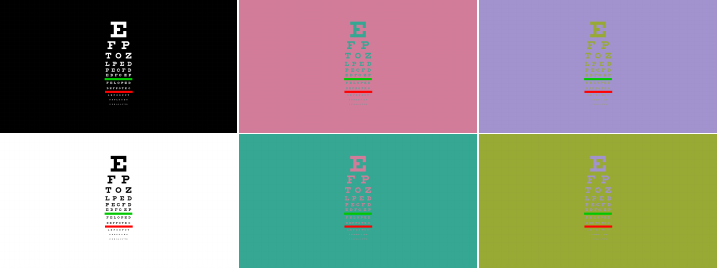} 
    \caption{Custom Snellen visual acuity charts used in the experiment.}    
    \label{fig:snellen-charts}
\end{figure}

Visual acuities for the observers in our experiment were estimated using customised Snellen charts shown in \figref{snellen-charts}. We used six different Snellen charts, corresponding to the different colour modulations: black-white, red-green, and yellow-violet. The charts were displayed on the same screen used in the main experiment, positioned at a fixed distance of 2.7\,m from the observer. The order of the charts' presentation was randomised for each observer. For each chart, observers were asked to read the letters on randomly chosen rows on the chart.  Rather than using the traditional line-by-line evaluation, where the observer must correctly identify all letters on a given line, we applied the more sensitive letter-by-letter scoring method as described in Monaco, Heimerl \& Kalb (2009)\cite{monaco2009clinically}. This method assigns partial credit based on the proportion of correctly identified letters within each line, which increases the precision of visual acuity measurement. The calculated Snellen fractions were converted to logMAR visual acuity values. 

\begin{figure}[!t]
    \centering
    \includegraphics[width=0.85\columnwidth]{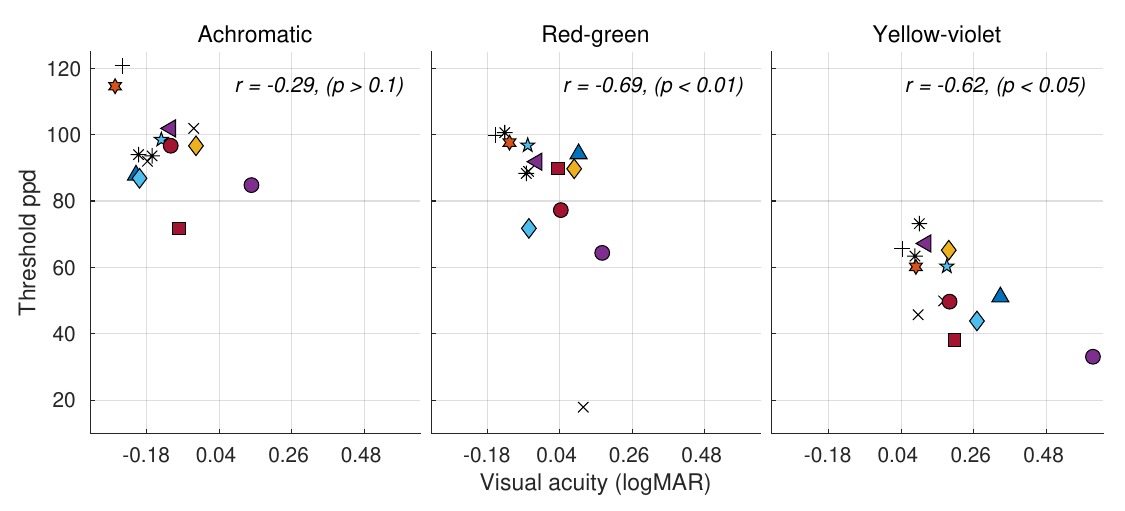}\\
    \setlength{\abovecaptionskip}{-2pt}
    \caption{Scatter plots showing the relationship between threshold pixels per degree (ppd) and visual acuity (logMAR) for individual observers. Spearman's rank correlation coefficient \textit{r} and the significance of the correlation are displayed on each plot. }
    \label{fig:min-ppd-va}
\end{figure}

For each colour channel (achromatic, red-green, yellow-violet), visual acuity measurements were averaged across the two opposite polarities. For instance, the achromatic visual acuity was calculated as the average of the measurements from both black-on-white and white-on-black Snellen charts. We then compared these averaged visual acuity values, expressed in logMAR, with the corresponding threshold ppd values derived from our experiment as shown in \figref{min-ppd-va}.  The relationship between these measures was assessed using Spearman’s rank correlation, with the correlation coefficient and significance level displayed for each colour modulation.

The results indicate that there is no statistically significant correlation between visual acuity and ppd thresholds for the achromatic condition, whereas statistically significant correlations were found for both isoluminant stimuli: red-green and yellow-violet. The low correlation in the achromatic condition may stem from the observers' familiarity with reading black-and-white text, which involves higher-level cognitive processes beyond visual resolution. This suggests that observers may use additional cues for identifying achromatic letters, making the achromatic letter identification task less dependent on low-level visual processing. On the other hand, the significant correlations observed for the isoluminant stimuli indicate that for purely colour-based contrasts, the resolution is more likely to be dictated by low-level visual processes. In these conditions, the visibility of letters appears to be closely tied to the actual resolution of the visual system, as there are fewer additional cues available to the observer. 

This finding could suggest the need for future investigations into whether visual acuities measured using letters accurately isolate low-level visual functions or if they are significantly influenced by higher-level contextual cues, such as familiarity with letter shapes and reading habits, and could be used to develop more precise methods of assessing visual acuity that reflect the fundamental capabilities of the visual system.

\subsection{Parameters of the resolution limit model}
\label{supp:model-fits}

\begin{figure}[!t]
    \centering
    \includegraphics[width=0.85\columnwidth]{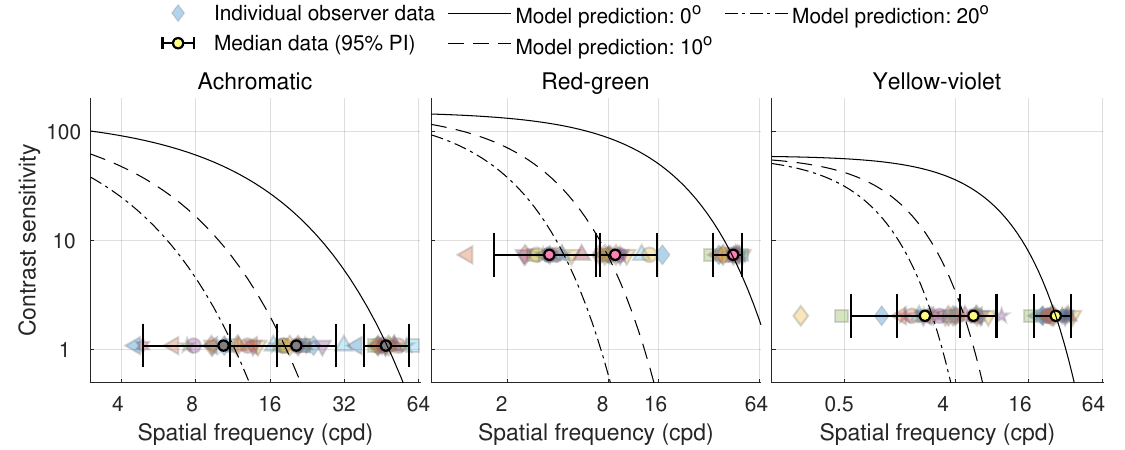}\\
    \caption{Contrast sensitivity functions fitted to the data using the Watson (2018)\cite{watson2018field} model. The model fits the parameters of the resulting contrast sensitivity function as functions of spatial frequency and eccentricity. In our experiment, the measured variable was the spatial frequency (in pixels per degree (ppd)) rather than the contrast (unlike conventional contrast sensitivity measurements) and the data points show the measured spatial frequency thresholds for fixed values of contrast. For an ideal display with no gamut limitations, the spatial frequency thresholds would be the values at sensitivity = 1.}
    \label{fig:min-ppd-csf}
\end{figure}

The parameters of the model in \eqref{res-limit-rho} were fitted using non-linear regression (\texttt{fitnlm} in MATLAB). The model was fitted separately for the three colour directions using the initial parameters reported in Watson (2018)\cite{watson2018field}. The fitted curve is shown as dashed lines in \figref{min-ppd-ecc}. The values of the estimated parameters along with the standard errors and 95\% confidence intervals are reported in \tableref{model_summary}. T-tests on parameter estimates indicated that all parameters were significantly different from zero ($p < 0.001$). For the assessment of the residuals' normality in our nonlinear model fitting, the Kolmogorov-Smirnov test was employed. For all three colour directions, the test did not reject the null hypothesis of normality with high p-values for achromatic ($p = 0.9456$), red-green ($p = 0.9046$), and yellow-violet ($p = 0.5977$) stimuli, suggesting that the residuals are normally distributed. \figref{min-ppd-csf}, visualises the resulting contrast sensitivity function (CSF) for different colour channels.

\begin{table}[!t]
\centering
\setlength{\abovecaptionskip}{5pt}
\caption{The estimated parameters of the fits for Watson (2018)\cite{watson2018field} model. Estimates are accompanied by their 95\% confidence intervals, standard errors, t-statistics, and corresponding p-values, indicating the significance of each parameter in the model. Results are shown separately for each of the three measured chromatic axes. }
\label{tab:model_summary}
\begin{tabular}{cccccc}
\toprule
\textbf{Parameter} & \textbf{Estimate} & \textbf{95\% Confidence Interval} & \textbf{Standard Error} & \textbf{t} & \textbf{p-value}\\
\midrule
\addlinespace
\multicolumn{6}{c}{\textit{Achromatic}} \\ 
$\log(\Sconscol{Ach})$ & 2.135 & [2.135, 2.135] & 4.004e-05 & 5.331e+04 & < 0.001 \\ 
$\krhocol{Ach}$ & -0.04394 & [-0.04778, -0.0401] & 0.00191 & -23.01 & < 0.001 \\ 
$\kecccol{Ach}$ & 0.1601 & [0.1294, 0.1908] & 0.01527 & 10.48 & < 0.001 \\ 
\midrule
\addlinespace
\multicolumn{6}{c}{\textit{Red-green}} \\ 
$\log(\Sconscol{RG})$ & 2.179 & [2.179, 2.179] & 2.609e-05 & 8.352e+04 & < 0.001 \\ 
$\krhocol{RG}$ & -0.02997 & [-0.03226, -0.02767] & 0.00114 & -26.28 & < 0.001 \\ 
$\kecccol{RG}$ & 0.4536 & [0.3823, 0.5248] & 0.0354 & 12.81 & < 0.001 \\ 
\midrule
\addlinespace
\multicolumn{6}{c}{\textit{Yellow-violet}} \\ 
$\log(\Sconscol{YV})$ & 1.777 & [1.776, 1.777] & 0.0001222 & 1.454e+04 & < 0.001 \\ 
$\krhocol{YV}$ & -0.05504 & [-0.06158, -0.04849] & 0.003252 & -16.92 & < 0.001 \\ 
$\kecccol{YV}$ & 0.4630 & [0.3462, 0.5798] & 0.05804 & 7.977 & < 0.001 \\ 
\bottomrule
\end{tabular}
\end{table}

\subsection{Probability distribution across the population}
\label{supp:prob-dist-model}

We model the probability distribution of spatial frequency thresholds (in pixels per degree, ppd) at various retinal eccentricities as normal distributions. These distributions are characterised by parameters, $\meangauss$ (mean), $\siggauss$ (standard deviation), and $\scgauss$ (scale). The random variable for these Gaussian distributions is a function of the spatial frequency, $f(\rho)$ as shown in \eqref{ppd-to-fppd}. The value of the scale parameter $\scgauss$ is fitted such that the integral of the Gaussian distribution, across the transformed spatial frequency $f(\rho)$, equals 1. The central value of the Gaussian ($\meangauss$) distribution at any given eccentricity is a function of the prediction from the model in \eqref{res-limit-rho}.

\begin{equation}
    G(\rho; \meangauss, \siggauss, \scgauss) = \frac{1}{\scgauss} \frac{1}{\siggauss \sqrt{2\pi}} \exp{\left(-\frac{(f(\rho)-\meangauss)^2}{2\siggauss^2}\right)}.
    \label{eq:gauss-dist} 
\end{equation}

The values of the parameters, $\siggauss$ and $\scgauss$, at the measured retinal eccentricities, are given in \tableref{gauss-params}. For eccentricities beyond our measured data points, we employ linear interpolation within the range of measured eccentricities, and linear extrapolation beyond the measured data points, to estimate these parameters. Specifically, we interpolate the standard deviation and scale using  MATLAB's \code{interp1} function with the 'linear' and 'extrap' options.

\begin{table}[!t]
  \centering
  \setlength{\abovecaptionskip}{5pt}
  \caption{The parameters $\siggauss$ and $\scgauss$ for different colour channels (achromatic, red-green, yellow-violet) at three levels of visual eccentricity (0, 10, 20 visual degrees).}
    \begin{tabular}{|l|l|c|c|}
    \hline
    \multicolumn{1}{|c|}{\textbf{Color direction}} & \multicolumn{1}{c|}{\textbf{Eccentricity}} & \multicolumn{1}{c|}{$\siggauss$} & \multicolumn{1}{c|}{$\scgauss$} \\
    \hline
    \multirow{3}{*}{Achromatic} & 0     & 0.1309 & 127.3 \\
    \cline{2-4} & 10    & 0.2203 & 176.0 \\
    \cline{2-4} & 20    & 0.2231 & 207.2 \\
    \hline
    \multirow{3}{*}{Red-green} & 0     & 0.1161 & 131.0 \\
    \cline{2-4} & 10    & 0.1452 & 232.7 \\
    \cline{2-4} & 20    & 0.1809 & 286.2 \\
    \hline
    \multirow{3}{*}{Yellow-violet} & 0     & 0.1729 & 154.9 \\
    \cline{2-4} & 10    & 0.2447 & 280.9 \\
    \cline{2-4} & 20    & 0.2069 & 343.6 \\
    \hline
    \end{tabular}%
  \label{tab:gauss-params}%
\end{table}%

\subsection{The effect of viewing distance}
\label{supp:view-distance}

We control the resolution of the stimulus in our experiment by adjusting the viewing distance. Yet, the viewing distance may have a sizeable effect on the detection of high-frequency patterns \cite{Depalma_Lowry_1962,Schober_Hilz_1965,Hernandez_Domenech_Segui_Illueca_1996}, with perceived resolution expected to be higher at larger distances. This is due to optical factors: diffraction and accommodative error. As the stimulus gets closer to our eyes, the pupil contracts (miosis). While the smaller pupil restricts the amount of light entering the eye and makes diffraction stronger, it also reduces optical aberrations, and improves depth-of-field at smaller distances, increasing the maximum perceivable resolution. Conversely, larger viewing distances reduce accommodation errors \cite{Charman_Radhakrishnan_2009} and, therefore, retinal blur, enhancing the ability to perceive high frequencies.

\begin{figure}[!b]
    \centering
    \includegraphics[width=\columnwidth]{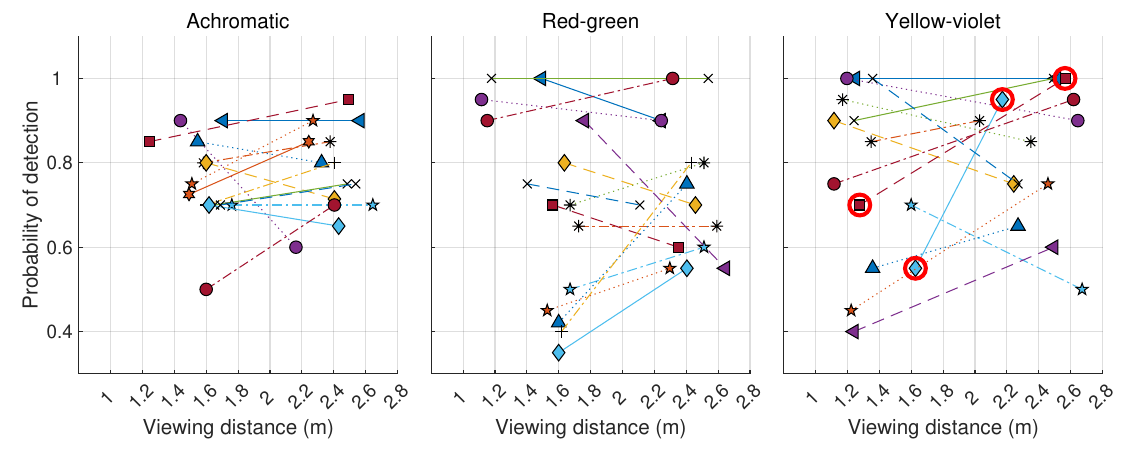}\\
    \setlength{\abovecaptionskip}{0pt}
    \caption{ Difference in the probability of detection between \textit{far} and \textit{near} stimuli. The figure shows individual observer data for achromatic, red-green, and yellow-violet stimuli at different viewing distances (m) for 16 observers. Each symbol represents data from a single observer. Negative slopes indicate higher detection probabilities for the near condition, while positive slopes indicate higher detection probabilities for the far condition. The data illustrate the variability in detection probability with changes in viewing distance, highlighting individual differences in visual sensitivity and the impact of viewing distance on the detection of spatial frequency across different colour channels. The data points with statistically significant differences between the two viewing conditions are highlighted with red circles.}
    \label{fig:min-ppd-vd-ind}
\end{figure}

We tested the effect of viewing distance in an additional experiment involving 16 observers. Stimuli with the same pixel-per-degree resolution were displayed at two different viewing distances, termed as \textit{far} and \textit{near}. The experiment was conducted only for the foveal condition, after the main experiment, to specifically test the effect of viewing distance on detection thresholds. \figref{min-ppd-vd-ind} shows the different probabilities of detection for stimuli shown at two viewing distances for different observers (estimated from 20 measurements). The viewing distances for each observer were chosen according to their threshold ppd values for the corresponding colour direction. To test the statistical significance of differences between the two viewing distances, we calculated the probabilities of difference between the two binomial distributions for \textit{near} and \textit{far} responses. To account for multiple comparisons, we applied the Holm-Bonferroni correction to the $p$-values from the two-tailed tests (at $\alpha=0.05$). After correction, we found that the differences in probabilities were statistically significant only for 2 observers and for the yellow-violet stimuli (see the circled data points in \figref{min-ppd-vd-ind}).

\begin{figure}[!t]
    \centering
    \includegraphics[width=0.4\columnwidth]{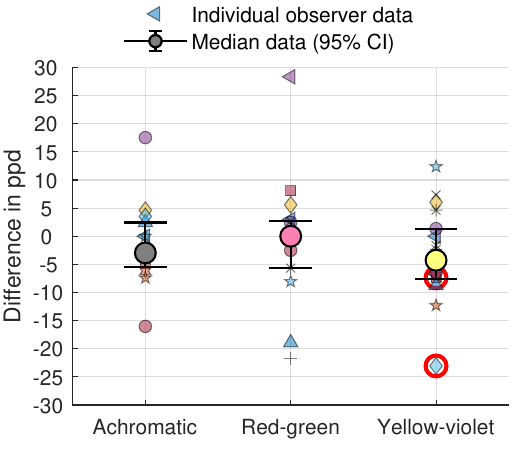}\\
    \setlength{\abovecaptionskip}{0pt}
    \caption{Difference in ppd thresholds between \textit{far} and \textit{near} stimuli. The figure shows individual observer data and median data with 95\% prediction intervals for achromatic, red-green, and yellow-violet stimuli for 16 observers. Negative values indicate higher thresholds for the near condition, while positive values indicate higher thresholds for the far condition. The red circles highlight individual data points, where the difference in binomial distributions of observer responses was found to be significant.}
    \label{fig:min-ppd-vd}
\end{figure}

To estimate the difference in ppd thresholds, we predicted the equivalent ppd thresholds for the \textit{near} and \textit{far} data from their probability of detection values using our psychometric function. \figref{min-ppd-vd} shows the differences in predicted thresholds between the two viewing distances. Negative values indicate higher thresholds for the \textit{near} conditions, and vice versa. For most observers, the differences between the two conditions were small within the range of our tested distances. The 2 yellow-violet differences identified as statistically significant are highlighted with red circles in \figref{min-ppd-vd}. Our data indicates that the effect of viewing distance on the resolution limit is small and inconsistent across the observers. Our protocol did not allow us to confirm statistical significance for the majority of the observers. For that reason, we do not account for the effect of viewing distance when reporting the results of our main experiment. 

\subsection{Foveated filtering}
\label{supp:img-sim}

\begin{figure}[!t]
    \centering
    \includegraphics[width=\textwidth]{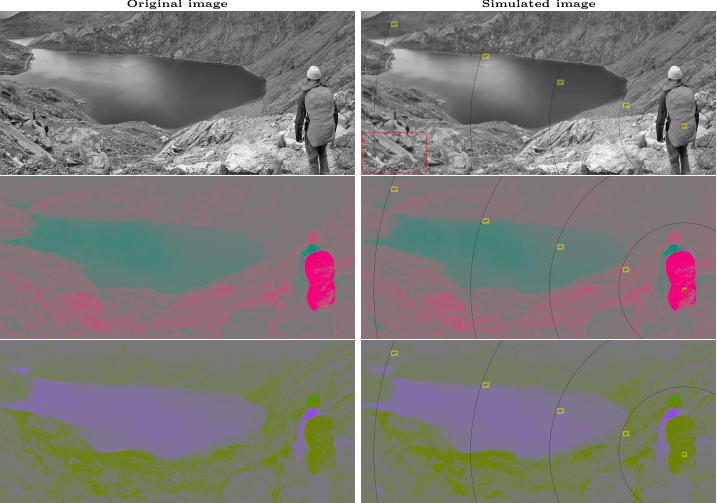} 
    \caption{Comparison of original and simulated images decomposed into three colour planes. The figure illustrates how foveated filtering selectively reduces spatial details based on retinal eccentricity, particularly in chromatic planes (red-green and yellow-violet). The simulated images (right) show diminished detail in peripheral areas, consistent with the filtering thresholds applied in the rendering process. To observe the intended effect, zoom into the figure such that the red outline in the top right image is approximately the size of a standard credit card.}
    \label{fig:graphics-appl-cols}
\end{figure}

Our foveated rendering application in Section \textit{Example: Foveated rendering} shows an example of eccentricity-dependent filtering on an image. The high-frequency details and the sub-threshold contrasts are selectively removed, based on retinal eccentricity, according to the predictions of our model. \figref{graphics-appl-cols} presents the decomposition of the original and processed images into their respective colour planes (achromatic, red-green, and yellow-violet). The original images on the left maintain full resolution across all colour channels, while the simulated images on the right show a progressive loss of detail, particularly in the red-green and yellow-violet planes.

\begin{figure}[!t]
    \centering
    \includegraphics[width=\linewidth]{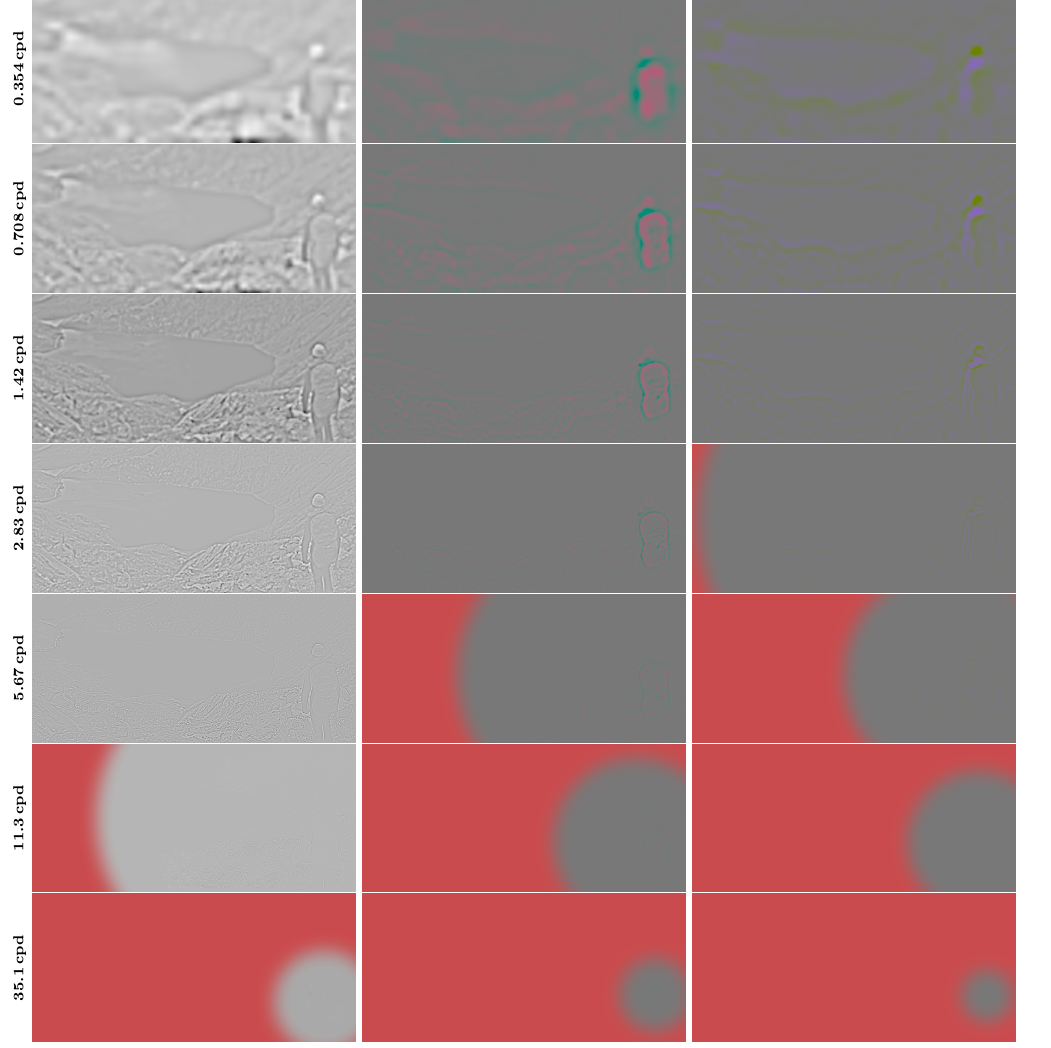} 
     \caption{Spatial frequency decomposition of the colour planes using a Laplacian pyramid coupled with the filtering mask. Each row corresponds to a different frequency band. The high-frequency details are progressively attenuated in the filtered image as a function of retinal eccentricity, preserving essential visual information while removing imperceptible details in the periphery. The red portions of the image show the areas where all the contrast information is removed.}
    \label{fig:img-lap-pyr}
\end{figure}

 \figref{img-lap-pyr} dissects the image processing by displaying the individual spatial frequency bands obtained from the Laplacian pyramid decomposition. Each row corresponds to a different frequency band, ranging from low to high spatial frequencies, filtered with an eccentricity-dependent mask. The progressive removal of high-frequency details, especially at larger eccentricities, is evident, particularly in the higher frequency bands (2.83 cpd and above).

\FloatBarrier

\FloatBarrier


\putbib[references]

\setcounter{page}{\value{savepage}}
\pagestyle{plain} 

\end{bibunit}

\end{document}